# Constraints from $^{26}$Al Measurements on the Galaxy's Recent Global Star Formation Rate and Core Collapse Supernovae Rate

F. X. Timmes[1,2], R. Diehl[1,3], and D. H. Hartmann[1,4]

[1] Department of Physics and Astronomy
Clemson University
Clemson, SC   29634;
fxt@burn.uchicago.edu, hartmann@grb.phys.clemson.edu.

[2] Astronomy & Astrophysics
University of California Santa Cruz
Santa Cruz, CA   95064;

[3] Max Planck Institute für Extraterrestrische Physik
85740 Garching, Germany;
rod@mpe-garching.mpg.de.

[4] Univeristäts-Sternwarte
37083 Göttingen, Germany

ABSTRACT

Gamma-rays from the decay of $^{26}$Al offer a stringent constraint on the Galaxy's global star formation rate over the past million years, supplementing other methods for quantifying the recent Galactic star formation rate, such as equivalent widths of H$\alpha$ emission. Advantages and disadvantages of using $^{26}$Al gamma-ray measurements as a tracer of the massive star formation rate are analyzed. Estimates of the Galactic $^{26}$Al mass derived from COMPTEL measurements are coupled with a simple, analytical model of the $^{26}$Al injection rate from massive stars and restrict the Galaxy's recent star formation rate to $5 \pm 4$ M$_\odot$ yr$^{-1}$. In addition, we show that the derived $^{26}$Al mass implies a present day Type II + Ib supernovae rate of $3.4 \pm 2.8$ per century, which seems consistent with other independent estimates of the Galactic core collapse supernova rate. If some independent measure of the massive star initial mass function or star formation rate or Type II + Ib supernovae rate were to become available (perhaps through estimates of the Galactic $^{60}$Fe mass), then a convenient way to restrain, or possibly determine, the other parameters is presented.

Subject headings: gamma rays: theory – nucleosynthesis, abundances – supernovae: general

## 1. INTRODUCTION

Through its 1.809 MeV gamma-ray line, Galactic $^{26}$Al was discovered in 1979 with the High Energy Astronomy Observatory-C spectrometer (Mahoney et al. 1982). Several measurements of the integrated 1.809 MeV flux have been performed since then, as reviewed by Prantzos & Diehl (1996). The most reliable of these measurements are probably derived from data obtained from the Gamma-Ray Spectrometer aboard the Solar Maximum Mission spacecraft (Share et al. 1985; Harris, Share, & Leising 1994), and the COMPTEL Imaging Telescope aboard the Compton Gamma-Ray Observatory (Diehl et al. 1995a). All estimates of the absolute $^{26}$Al mass in the Galaxy rest on assumptions about the spatial distribution of the sources, as the 1.809 MeV measurements themselves do not carry distance information. This situation may change with future high-resolution, high-sensitivity instruments if the line shape and Doppler shift of the 1.809 MeV line can be extracted (Gehrels & Chen 1996).

From the HEAO-C data, Mahoney et al. determined 3 $M_\odot$ of $^{26}$Al, assuming the smooth spatial distribution derived from COS-B measurements of Galactic gamma-rays in the 100 MeV regime. All other non-imaging instruments basically confirm this result, based on the same (or equivalent) assumptions about spatial source distributions. It became evident, however, with the COMPTEL imaging data ($\simeq 5°$ degree spatial resolution) that the distribution of 1.809 MeV emission is significantly different than that derived from COS-B measurements (Diehl et al. 1995a,b). Although the ridge of the Galactic plane dominates the emission, several prominent regions of emission (particularly Cygnus and Vela), suggest substantial deviations from a smooth emission pattern following the distribution of gas. The COMPTEL team fitted several models of candidate source distributions to the COMPTEL data, such as CO survey data (Dame et al. 1987), analytical models based on exponential disks, and H II region data evaluated in the context of spiral-arm structure (Taylor & Cordes 1993). If the prominent, localized regions of emission beyond the inner Galaxy are excluded from a fit to such models, then all fits to axisymmetric models yield a Galactic mass of 3 ± 0.5 $M_\odot$ (Diehl et al. 1995; Knödlseder et al. 1996). However, the marked asymmetry in the emission profile along the disk suggests that spiral structure is important (Prantzos & Diehl 1996; Chen, Gehrels, & Diehl 1995). If one adopts a composite model of disk-like emission plus contributions from sources along spiral-arms, then fits to the COMPTEL data results in a total Galactic mass of 2.7 ± 0.8 $M_\odot$, with 0.7 ± 0.3 $M_\odot$ attributed to the spiral arm component (Knödlseder et al. 1996).

Candidates for the origin of $^{26}$Al include massive stars (through their supernovae ejecta and Wolf-Rayet wind phase contributions), asymptotic giant branch (AGB) stars, and classical novae with metal enriched atmospheres (see Prantzos & Diehl 1996 for details) and While yields from all these candidate $^{26}$Al sources are uncertain, it seems safe to assume that core-collapse supernovae as the dominant source will not be challenged. Thus, the COMPTEL measurements and analysis can be interpreted as: 0.7 M$_\odot$ of $^{26}$Al if only the spiral-arm component is assigned to Type II + Ib supernovae, to 2.5 M$_\odot$ of $^{26}$Al if $\simeq 80\%$ of the emission is assigned to Type II/Ib supernovae and no foreground contributions from localized emission regions similar to Vela or Cygnus lie in the direction of the inner Galaxy. Since the decay time of $^{26}$Al ($\tau_{1/2} = 7.5 \times 10^5$ yr) is short compared to Galactic rotation timescales ($\tau_{\rm Gal} \simeq 10^8$ yr) this estimated $^{26}$Al mass range serves as an important constraint of the stellar population responsible (i.e., massive stars) for the synthesis.

One of the recurring concepts of this paper is the relationship between the derived $^{26}$Al mass estimates, initial mass function (henceforth IMF), recent global star formation rate (henceforth SFR), and present epoch Type II + Ib supernovae rate. The connections made in this paper between these four quantities are new, or at least not widely recognized, but the raw science of the four quantities themselves is not new. It is worthwhile in making the connections to succinctly summarize the shape of the IMF, tracers of the Galactic SFR, and stellar core-collapse estimates.

## 2. STAR FORMATION RATE CONSTRAINTS

### 2.1. *The Initial Mass Function*

The first empirical determination of the observed IMF showed that the number of stars between 0.4 M$_\odot$ and 10 M$_\odot$ could be described as a power-law with the index $\gamma = -2.35$ (Salpeter 1955). Studies since then (Miller & Scalo 1979; Humphreys & McElroy 1984; Scalo 1986; Rana 1991; Tinney, Mould & Reid 1992; Parker & Garmany 1993; Reid 1994; Hunter 1995; Massey et al. 1995a,b; Kroupa 1995; Méra, Chabrier & Baraffe 1996; Mayya & Prabhu 1996; Hunter et al. 1996) suggest that the observed IMF becomes flatter than a pure power-law at the smallest stellar masses ($\gamma \sim -1$ for m $\leq 0.5$ M$_\odot$) and becomes steeper for the most massive stars ($\gamma \sim -3.3$ for m $> 10$ M$_\odot$). Some studies indicate that the IMF has more structure than either a power-law or log-normal form (Rana 1991), while others argue that the IMF is closer to a power-law and hence has less structure (Scalo 1986). Overall, the shape of the IMF appears to be quite robust (centered on the Salpeter $-2.35$ exponent) and seems not to change very much from one star-forming region to another. In any event, proceeding from the observed luminosity function to the implied IMF depends upon the stellar evolutionary tracks used in the fitting procedure (Tinsley 1980; Elmgreen 1995a,b; Efremov 1995; Adams & Fatuzzo 1996; Arnett 1996).

### 2.2. *Galactic Star Formation Rates*

Despite many uncertainties, the results of various studies (Schmidt 1959, 1963; Searle, Sargent & Bagnuolo 1973; Larson & Tinsley 1974; Cohen 1976; Huchra 1977; Smith, Biermann & Mezger 1978; Lequeux 1979; Talbot 1980; Tinsley 1980; Kennicut 1983; Kennicut & Kent 1983; Güsten & Mezger 1983; Turner 1984; Lacey & Fall 1985; Dopita 1985; Gallagher, Bushouse, & Hunter 1989; Romanishin 1990; Rana 1991; Kennicut 1992; Lada 1992; Gallagher & Gibson 1993; Kennicut, Tamblyn & Congdon 1994; Hill, Madore, & Freedman 1994; Gallager & Scowen 1995; Gallego et al. 1995; Lada & Lada 1995) lead to the picture of $\sim 10\%$ of the current SFR occuring in the innermost 1 kpc of the Galaxy, and most of the remaining 90% concentrated between 5 and 9 kpc from the center, which is where most of the Galaxy's giant molecular clouds, infrared emission, and other signs of intense star formation reside. The current SFR for the whole Galaxy has been estimated to be 0.8 $M_\odot$ yr$^{-1}$ (Talbot 1980), 3.0 $M_\odot$ yr$^{-1}$ (Turner 1984), 5.3 $M_\odot$ yr$^{-1}$ (Smith et al. 1978), 13.0 $M_\odot$ yr$^{-1}$ (Güsten & Mezger 1982), and 6.0 $M_\odot$ yr$^{-1}$ (Pagel 1994). Güsten & Mezger (1982) also estimated the massive star SFR in the spiral-arms to be $5 \pm 2$ $M_\odot$ yr$^{-1}$. Note that all of these estimates use various indicators of the massive star population, and then convert these indicators to a total SFR by means of an assumed (universal) IMF.

### 2.3. H$\alpha$ Line Widths

Equivalent widths of H$\alpha$ emission have been the best available and most popular method for quantifying the present SFR. This is because H$\alpha$ equivalent widths are directly proportional to the number of Lyman continuum photons emitted by massive stars and hence proportional to the SFR. Other SFR measures or indicators (H$\beta$, H$\gamma$, [O III] $\lambda$5007, [O II] $\lambda$3727, integrated UBV colors, infrared luminosities, IRAS fluxes, free-free radio emission from H II regions, magnetic field strengths, and brightest individual star counts) are more affected by stellar absorption, interstellar reddening, excitation strength, metallicity, dust abundances, dust composition, incompleteness, sky coverage, or resolution limitations than H$\alpha$ emission (see references above). Even when H$\alpha$ measurements are combined with some of the alternative indicators, the derived SFR is a lower limit since even H$\alpha$ is not completely immune (just less sensitive) from the contaminants listed above. It has been suggested that near-infrared recombination lines of Br$\gamma$ could be an even better measure of the current SFR (Leitherer & Heckmann 1995), but instrumentation difficulties impede progress along this avenue at present.

### 2.4. Gamma-ray Measurements

Gamma-rays from the decay of $^{26}$Al offer a unique measure of the present SFR in the Galaxy. Several of the difficulties noted above are mitigated by the transparency of the Galaxy to gamma-rays (e.g., absence of interstellar reddening), but several difficulties remain (e.g., spatial resolution limitations). Nevertheless, gamma-rays offer a complimentary indicator of the Galaxy's present epoch SFR.

The IMF by number (assumed universal and constant independent), the normalization condition, and the normalization constant are

$$f(m) = A \, m^\gamma$$
$$\int_{M_L}^{M_U} f(m) \, dm = 1 \tag{1}$$
$$A = (\gamma + 1) \left( M_U^{\gamma+1} - M_L^{\gamma+1} \right)^{-1} \qquad \gamma \neq -1 \quad ,$$

respectively. The mean mass of stars, the fraction of all stars which become core-collapse supernovae (Type II + Type Ib), and the steady-state core-collapse supernovae rate, respectively, are

$$\begin{aligned} <m> &= \int_{M_L}^{M_U} f(m) \, m \, dm = \frac{A}{\gamma+2} \left( M_U^{\gamma+2} - M_L^{\gamma+2} \right) \qquad \{M_\odot\} \\ F_{SN} &= \int_{M_{SN}}^{M_U} f(m) \, dm = \left[ 1 - \left( \frac{M_{SN}}{M_U} \right)^{\gamma+1} \right] \left[ 1 - \left( \frac{M_L}{M_U} \right)^{\gamma+1} \right]^{-1} \\ R_{SN} &= \dot{N}_* \, F_{SN} = \frac{\Psi}{<m>} F_{SN} \qquad \left\{ \frac{\text{number}}{\text{year}} \right\} , \end{aligned} \tag{2}$$

where $\dot{N}_*$ is the stellar birthrate in number per year, $\Psi$ is the SFR in $M_\odot$ per year, $M_L$ is the smallest stellar mass in the distribution, $M_U$ is the largest stellar mass in the distribution, and $M_{SN}$ is the smallest stellar mass which undergoes core-collapse. The mean yield of $^{26}$Al from Type II + Ib events is

$$\begin{aligned} <y> &= \int_{M_{SN}}^{M_U} f(m) \, y(m) \, dm \, \left[ \int_{M_{SN}}^{M_U} f(m) \, dm \right]^{-1} \\ &= \frac{1}{F_{SN}} \int_{M_{SN}}^{M_U} f(m) \, y(m) \, dm \\ &= \frac{y_{\text{eff}}}{F_{SN}} \qquad \{M_\odot\} \, . \end{aligned} \tag{3}$$

Finally, the steady state injection rate of $^{26}$Al is

$$\dot{M}_{26} = R_{SN} \, <y> = \frac{\Psi \, y_{\text{eff}}}{<m>} \qquad \left\{ \frac{M_\odot}{\text{year}} \right\} . \tag{4}$$

Eq. (4) may be solved for the global SFR, $\Psi$, for a given observed $^{26}$Al mass in a steady state galaxy, and a given IMF exponent $\gamma$. The results of such a procedure is shown in the lower panel of Figure 1. Each labeled curve corresponds to a different Galactic $^{26}$Al mass (in solar masses), with the preferred $^{26}$Al mass range (0.7 - 2.8 $M_\odot$) imposed by the COMPTEL observations shown as the grey band. Integration limits of $M_L = 0.1 \, M_\odot$, $M_U = 40 \, M_\odot$, and $M_{SN} = 10 \, M_\odot$ were used in constructing Fig. 1, but the chief conclusions are quite robust with respect to reasonable variations in the integration limits. The mean and effective $^{26}$Al yields in eqs. (3) and (4) were calculated with the $^{26}$Al mass ejected in the Woosley & Weaver (1995) massive star models. There is about an order of magnitude difference in the $^{26}$Al yields if the results of the Thielemann, Nomoto, & Hashimoto (1996) survey are used instead of Woosley & Weaver. The bulk of the synthesis of this radioactive isotope takes place in the presupernova star. It is imperative to follow this stage of the star's evolution with a sufficient nuclear reaction network, especially during the last few hours of convective neon and oxygen burning. Woosley & Weaver used a 200 isotope network from the main sequence through the explosion, while Thielemann et al. follow the presupernova evolution from an initial helium core mass with an $\alpha$-chain network. Only during the explosive phases of the evolution do Thielemann et al. switch to a larger reaction network. This accounts for most of the difference in the $^{26}$Al production in the two surveys. Deviation from straight lines in the lower panel of Fig. 1 is due to the IMF exponent approaching the removable singularity at $\gamma = -1$ (see eq. 1). Only a mathematical reason, not a physical one, is responsible for the flattening of the curves.

The horizontal dimension of the dashed box in the lower panel of Fig. 1 is centered on the Salpeter $-2.35$ exponent, and is representative of the range of IMF exponents for massive stars encountered in the literature. Vertical dimensions of the dashed box were set by requiring consistency between the COMPTEL estimates of the Galactic $^{26}$Al mass and the simple model for the $^{26}$Al injection rate from massive stars. The lower panel of Fig. 1 suggests that the global SFR in the Galaxy during the past million years is restricted to $5 \pm 4$ M$_\odot$ yr$^{-1}$. This is consistent with the Güsten & Mezger (1982) H$\alpha$ estimate of the massive star SFR in the spiral arms of $5 \pm 2$ M$_\odot$ yr$^{-1}$, and the more recent determinations of the Galaxy's global SFR (see §1).

Eq. (2) may be solved for the core-collapse supernova rate given the global SFR $\Psi$ and the IMF exponent $\gamma$. This solution is shown in the upper panel of Fig. 1, for the SFRs calculated in constructing the lower panel of Fig. 1. As before each labeled curve corresponds to a different Galactic $^{26}$Al mass, with the preferred $^{26}$Al mass range $(0.7 - 2.8$ M$_\odot)$ imposed by the COMPTEL observations shown as the grey band. Here the curves are straight lines (as expected) since the approach to the removable singularity at $\gamma = -1$ is embedded in both factors ($<m>$, $F_{SN}$) of eq. (2) and they cancel each other. For the same plausible range of IMF exponents considered above, the COMPTEL estimates of the Galactic $^{26}$Al mass appear to imply a core-collapse supernovae rate of $3.4 \pm 2.8$ per century.

### 3. DISCUSSION

Direct measurement of the Galactic supernova rate is difficult owing to possible incompleteness in historical observations, and uncertainty as to the fraction of the Galactic disk and altitude that are sampled. Indirect inference from supernova rates in similar galaxies is adversely affected by the imprecise value of the Hubble constant, and the uncertainty in estimating the total blue luminosity and morphological classification of our Galaxy. Systematic searches for extragalactic supernova are also hampered by the need to know the distance, luminosity, and Hubble class of the host galaxy, as well as the dates and limiting magnitude of each observation. Such detailed information is only available in a few dozen supernova catalogs. Based on these surveys, estimates of the core-collapse and thermonuclear driven supernova rates were derived and discussed by van den Bergh & Tammann (1991), and Cappellaro (1993). These estimates assumed that the peak luminosity of each supernova class was a standard candle, and a large correction for edge-on spirals ($\sin i$ effect). Using the extragalactic estimates with a total Galactic blue luminosity of $2.3 \times 10^{10}$ $L_\odot$, a Hubble constant of 75 km s$^{-1}$ Mpc$^{-1}$, and a Sbc Galactic morphology, the Galactic core-collapse supernova rate has been estimated to be 4.1 per century (van den Bergh & Tammann 1991) and $2.4 - 2.7$ per century (van den Bergh & McClure 1994; Tamman, Löffler, & Schröder 1994). These estimates agree (perhaps auspiciously) with the core-collapse supernovae rate implied by a near-Salpeter IMF exponent and the COMPTEL derived $^{26}$Al mass.

While the general agreement found between estimates of the COMPTEL derived $^{26}$Al mass, the range of massive star IMF exponents encountered in the literature, complimentary measures of the recent SFR, and the present epoch Type II + Ib supernovae rate may be fortuitous and reminiscent of epicycles, it does point to a consistent picture. Less speculative is the fact that gamma-rays from the decay of certain radioactive nuclei, such as $^{26}$Al and $^{60}$Fe, offer a unique measure of the present SFR in the Galaxy that is complimentary to other popular indicators of the Galaxy's present epoch SFR (e.g, H$\alpha$, H$\beta$, H$\gamma$, [O III] $\lambda$5007, [O II] $\lambda$3727, integrated UBV colors, infrared and radio luminosities, and stellar counts). If some independent measure of the massive star IMF exponent or SFR or Type II + Ib supernovae rate were to become available (perhaps through measurements of the Galactic $^{60}$Fe mass), then Fig. 1 offers a convenient way to constrain, or possibly even determine, the other parameters.

This work was done at Clemson University and supported by NASA grant NAG 5-1578 (D.H.H), a Godfrey International Scholar Fellowship (R.D), the Keck Foundation (R.D), and a Compton Gamma Ray Observatory Postdoctoral Fellowship (F.X.T). D.H.H thanks the Academy of Sciences at Göttingen University for support through the C. F. Gauss endowment. We thank John Cowan for a pertinent referee report.

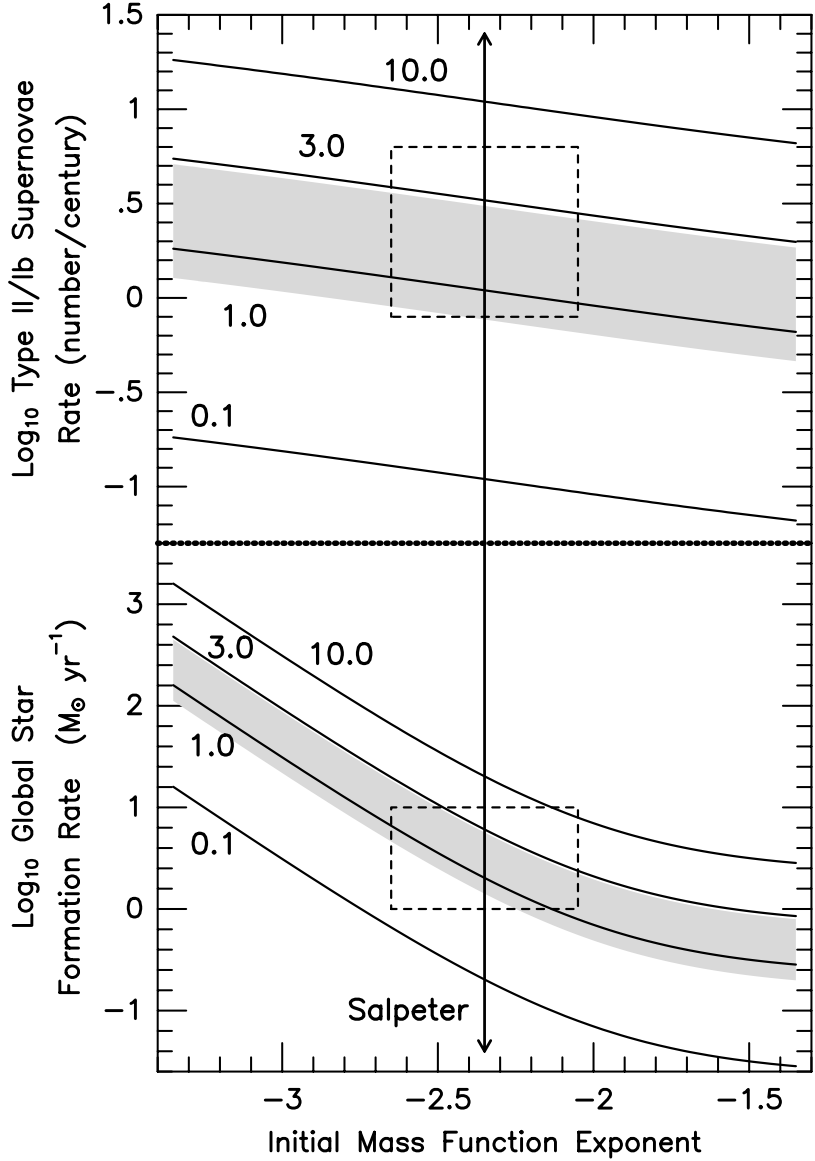

Fig. 1.— Global star formation rate (lower panel) and current Galactic Type II + Ib supernova rate (upper panel) vs the IMF exponent. Each labeled curve corresponds to a different Galactic $^{26}$Al mass (in solar masses), with the preferred $^{26}$Al mass range (0.7−2.8 M$_\odot$) imposed by the COMPTEL observations shown as the grey bands. The horizontal dimension of the dashed boxes are centered on a Salpeter −2.35 exponent, and are representative of the range of IMF exponents for massive stars encountered in the literature. Vertical dimensions of the dashed boxes were set by requiring consistency between the COMPTEL estimates of the Galactic $^{26}$Al mass and the simple model for the $^{26}$Al injection rate from massive stars. This consistency then appears to imply a Galactic SFR during the past million years of 5 ± 4 M$_\odot$ yr$^{-1}$, and a core-collapse supernovae rate of 3.4 ± 2.8 per century.